\documentclass[
  twocolumn,
  prl,
  amsmath,
  amssymb,
  superscriptaddress,
  nofootinbib,
  floatfix
]{revtex4}

\usepackage{mathtools}
\usepackage{bbm}
\usepackage{bm}
\usepackage{dsfont}
\usepackage{graphicx}
\usepackage{pifont}
\usepackage{textcomp}
\usepackage{gensymb}
\usepackage{accents}
\usepackage{upgreek}
\usepackage{array}
\usepackage{hhline}
\usepackage{tabularx}

\makeatletter
\def\NAT@bibsetnum#1{%
 \setlength{\topsep}{\z@}%
 \NATx@bibsetnum{#1}%
}%
\renewenvironment{thebibliography}[1]{%
 \NAT@thebibliography{#1}%
 \@clubpenalty\clubpenalty
 \let\@TBN@opr\present@bibnote
 \@FMN@list
}{%
 \@endnotesinbib
 \edef\@currentlabel{\arabic{NAT@ctr}}%
 \NAT@endthebibliography
 \global\let\auto@bib\@empty   
}
\newcommand*{\supplementarystart}{%
  \close@column@grid%
  \clearpage%
  \onecolumngrid%
  \setcounter{enumiv}{0} 
  \setcounter{equation}{0} 
  \setcounter{figure}{0} 
  \setcounter{table}{0} 
  \setcounter{page}{1}
  \c@secnumdepth=4
  \renewcommand{\theequation}{S\arabic{equation}} 
  \renewcommand{\bibnumfmt}[1]{[s##1]} 
  \renewcommand{\@onlinecite}{s\citealp} 
  \renewcommand{\cite}[1]{{[}\onlinecite{##1}{]}}
  \renewcommand{\thefigure}{S\arabic{figure}}
  \renewcommand{\thetable}{S\Roman{table}}
  \renewcommand{\thepage}{S\arabic{page}}
}
\makeatother

\DeclareMathOperator{\str}{str}
\DeclareMathOperator{\Str}{Str}

\DeclareMathOperator{\tr}{tr}

\DeclareMathOperator{\diag}{diag}

\newcommand{\be}{\begin{equation}}
\newcommand{\ee}{\end{equation}}
\newcommand{\bi}{\begin{itemize}}
\newcommand{\ei}{\end{itemize}}
\newcommand{\bea}{\begin{eqnarray}}
\newcommand{\eea}{\end{eqnarray}}
\newcommand{\bem}{\begin{multline}}
\newcommand{\eem}{\end{multline}}
\newcommand{\Hc}{\mathcal{H}}
\newcommand{\U}{{\rm U}}

\begin{document}


\title{Localization effects on magnetotransport of a disordered Weyl semimetal}
\author{E.\ Khalaf}
\affiliation{Max Planck Institute for Solid State Research, Heisenbergstr.\ 1, 70569 Stuttgart, Germany}
\author{P.\ M.\ Ostrovsky}
\affiliation{Max Planck Institute for Solid State Research, Heisenbergstr.\ 1, 70569 Stuttgart, Germany}
\affiliation{L.\ D.\ Landau Institute for Theoretical Physics, 142432 Chernogolovka, Russia}

\begin{abstract}
We study magnetotransport in a disordered Weyl semimetal taking into account localization effects exactly. In the vicinity of a Weyl node, a single chiral Landau level coexists with a number of conventional non-chiral levels. Disorder scattering mixes these topologically different modes leading to peculiar localization effects. We derive the average conductance as well as the full distribution function of transmission probabilities along the field direction. Remarkably, we find that localization of the non-chiral modes is greatly enhanced in a strong magnetic field with the typical localization length scaling as $1/B$. Technically, we use the non-linear sigma-model formalism with a topological term describing the chiral states. The problem is solved exactly by mapping to an equivalent transfer matrix Hamiltonian. 
\end{abstract}

\maketitle

\emph{Introduction.---}
Weyl semimetals have received considerable interest in the past years due to their unusual transport properties and exotic surface states \cite{Wan11, Hosur13, Burkov15, Kim13, Xu15, Xu15b, Huang15, Huang15b, Shekhar15, Lv15, Yang15, Xiong15, Zhang16}. A Weyl semimetal (WSM) is a three-dimensional (3D) analog of graphene characterized by the existence of isolated touching points between valence and conduction bands with linear electron dispersion. Each band touching point has a definite chirality and can be viewed as a magnetic monopole (source or sink of Berry flux) in momentum space. The topological nature of such Weyl points protects their linear spectrum from gap opening and significantly alters  transport properties of the material compared to usual metals.

\begin{figure}
\center
\includegraphics[width=0.36\textwidth]{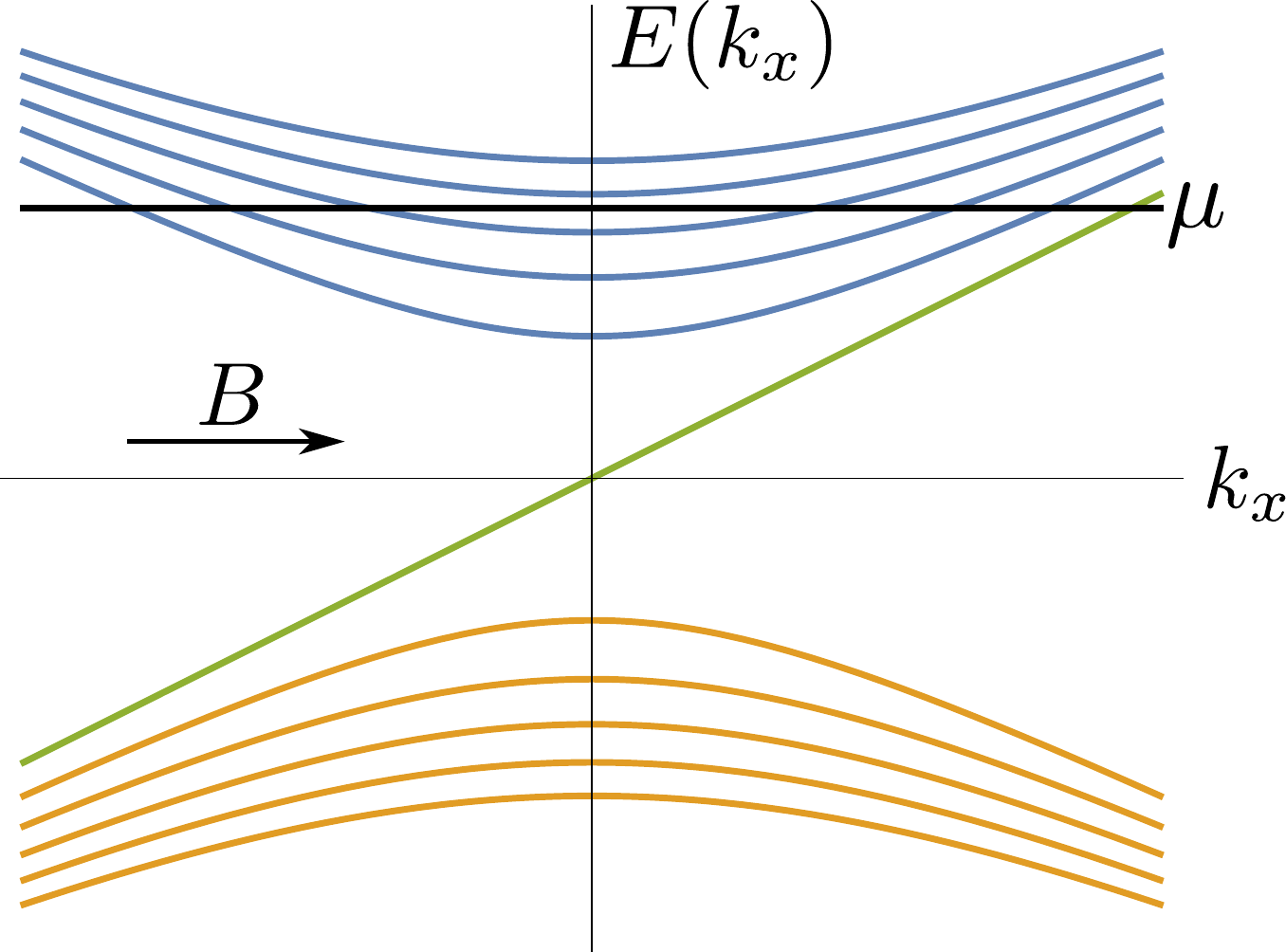}
\caption{The Landau levels in the vicinity of a Weyl point in magnetic field $B$. The levels are non-dispersive in the plane transverse to the magnetic field. The energy-momentum dependence along the field (\protect\ref{Landau}) is displayed. Each level has a macroscopic degeneracy $m = BA/\Phi_0$. The system is effectively a quasi-1D wire with both chiral and non-chiral channels provided $L \gg \sqrt{A}$.}
\label{Weylf}
\end{figure}

One immediate manifestation of the topological nature of the WSM spectrum is provided by the structure of Landau levels in an external magnetic field. Electron motion is confined in the plane perpendicular to the field yielding discrete energy levels
\begin{equation}
 E_n
  = \hbar v \begin{cases}
      \mathop{\mathrm{sign}} n \sqrt{4 \pi |n B|/\Phi_0 + k_x^2}, & n \neq 0, \\
      k_x, & n = 0.
    \end{cases}
 \label{Landau}
\end{equation}
Here, $v$ is the Fermi velocity characterizing linear dispersion at the Weyl node and $\Phi_0 = hc/e$ is the magnetic flux quantum. Each discrete level has a macroscopic degeneracy $m = B A/\Phi_0$, where $A$ is the total sample cross-section area. Landau level dispersion is illustrated in Fig.\ \ref{Weylf}. The topological property of the Weyl point is manifested by a single unidirectional (chiral) level with $n = 0$, in addition to many non-chiral levels ($n \neq 0$). A chiral level propagating in the opposite direction belongs to another Weyl node and is well separated in the momentum space. This results in a very large negative magnetoresistance in the direction parallel to the field \cite{Nielsen83, Fukushima08, Zyuzin12, Aji12, Xiong15}.

Despite the number of theoretical works discussing transport in WSM \cite{Hosur12, Son13, Parameswaran14, Burkov14, Gorbar14,Sbierski14, Burkov15, Klier15, Altland16,Baireuther16}, effects of disorder have not been taken into account beyond the semiclassical weak localization regime so far. This is justified in the absence of magnetic field since the material is three-dimensional and localization effects are weak. For magnetotransport, the situation is drastically different. Since Landau levels disperse only along the field, electron transport is effectively quasi-one-dimensional (1D) \cite{Altland16} and hence subject to strong localization effects.

Scattering on impurities eventually localizes the non-chiral modes of the spectrum, while the chiral level evades localization. This protected chiral Landau level is the primary source of strong magnetoresistance. Backscattering of chiral states is only possible when impurities couple different Weyl nodes of opposite chirality, which occurs at parametrically larger length scales compared to the mean free path within each node.   

In this Letter, we study transport in a generic quasi-1D system which hosts both chiral and non-chiral modes fully taking into account localization effects. Scattering on impurities eventually localizes the non-chiral modes, while the chiral level evades localization. This model describes longitudinal magnetotransport in a WSM at finite doping with some fixed nonzero chemical potential (cf.\ Fig.\ \ref{Weylf}). For a WSM sample with macroscopic length $L$ and cross-section $A$, quasi-1D geometry means $L \gg \sqrt{A}$. The surface states are ignored since their contribution to transport is negligible in a bulk sample \cite{footnote1}.

We will show that, in the presence of $m$ chiral channels, typical localization length for non-chiral modes is significantly reduced, $\xi^\text{typ} = \xi/(1 + m)$. Here $\xi = \sigma A$ is the localization length the system would have in the absence of chiral modes, $\sigma$ is the 3D conductivity (in units $e^2/h$ per Weyl node) given by the product of electron charge, density of states, and diffusion constant $D = v l$. For magnetotransport in WSM with a macroscopic degeneracy of the chiral Landau level $m = B A/\Phi_0\gg 1$, typical localization length for non-chiral modes $\xi^\text{typ} = \sigma \Phi_0/B$ is independent of $A$. This effectively 1D localization effect is so strong that it can be observed in a WSM sample of arbitrary thickness. The enhancement of localization of the non-chiral channels can be understood in terms of statistical level repulsion between the transmission eigenvalues in the system, where the presence of chiral channels with perfect transmission suppresses transport from the remaining non-chiral ones.

Aside from magnetotransport in a WSM, quasi-1D wire model with chiral channels applies to a number of other systems such as the interface between two quantum Hall samples \cite{Grayson05, Grayson07, Grayson08, Steinke08, Khalaf16} and graphene zigzag nanoribbons \cite{Wakabayashi07}. We obtain closed analytic expressions for the conductance as a function of $L$, disorder strength, and number of chiral channels taking into account localization effects exactly. In addition, we also consider the distribution function of transmission eigenvalues \cite{Dorokhov84, Nazarov94}, that contains information about shot noise and other higher moments of full-counting statistics \cite{Nazarov94, Lee95, Nazarov09}.

\emph{Formalism.---}
We consider a model of a quasi-1D wire of length $L$ attached to two leads with $N \gg 1$ channels in total and imbalance $m$ (difference of right- and left-propagating modes). Disorder mixes all channels at the scale of the mean free path $l$. The full information about electron transport is captured by the distribution function of $N$ transmission probabilities which we will explicitly derive.

The main technical tool we use to describe the disordered system is the supermatrix non-linear sigma model \cite{Efetov83, Efetov99, Mirlin00, Khalaf16}. It can be derived following the conventional procedure \cite{Efetov83, Efetov99, Mirlin00} starting from the effective action for the non-interacting electrons, averaging over disorder potential, decoupling of the fields with the help of Hubbard-Stratonovich transformation (that introduces the matrix $Q$). Integrating out the original electron fields, and performing the gradient expansion at the saddle-point manifold yields the sigma model action \cite{Khalaf16, Altland16}
\begin{equation}
 \begin{gathered}
  S[Q]
   = -\int_0^L \! \! {\rm dx} \str \left[
       \frac{\xi}{8} \left( \frac{\partial Q}{\partial x} \right)^2
       + \frac{m}{2} T^{-1} \Lambda \frac{\partial T}{\partial x}
     \right], \\
  \xi
   = N l = \sigma A,
  \quad
  Q
   = T^{-1}\Lambda T,
  \quad
  \Lambda
   = \mathop{\mathrm{diag}} (1,-1)_\textrm{RA}.
 \end{gathered}
 \label{Snlsm}
\end{equation}
The field $Q$ is a $4 \times 4$ supermatrix operating in the product of retarded-advanced (RA) space and the Bose-Fermi superspace (BF). The supertrace ``str'' is defined by $\str A = \tr A_\text{BB} - \tr A_\text{FF}$ as in Ref.\ \cite{Mirlin00}.

The sigma model (\ref{Snlsm}) describes magnetotrancport in a WSM in the vicinity of a single Weyl node. Effect of the coupling between different nodes will be discussed later.

The matrix $Q$ obeys the nonlinear constraint $Q^2 = 1$ and can be parametrized in terms of the unitary supermatrix $T$ as indicated in Eq.\ (\ref{Snlsm}). The matrix $Q$ is invariant under the gauge transformation $T \mapsto KT$ for any matrix $K$ that commutes with $\Lambda$. However, the second term of the action (\ref{Snlsm}) is written as a functional of $T$ rather than $Q$ and changes by an integral of the total derivative $(m/2) \partial_x \str (\Lambda \ln K)$ under such a transformation. The reason for this is that the theory (\ref{Snlsm}) describes a single Weyl node. The gauge invariance is restored when a Weyl node of opposite chirality is included. It is worth noting that this second term naturally appears in the field theory of a quantum Hall edge \cite{Pruisken84} and constitutes a 1D version of the WZW term \cite{Wess71, Witten83, Witten84}.

In order to access transport characteristics of the system, we apply twisted boundary conditions with the counting field introduced by Nazarov \cite{Nazarov94, Rejaei96, Khalaf16}
\begin{equation}
\label{BC}
 Q(0)
  = \Lambda,
 \quad
 Q(L)
  = \begin{pmatrix}
     \cos \hat{\theta} & \sin \hat{\theta} \\ \sin \hat{\theta} & -\cos \hat{\theta}
    \end{pmatrix}_\mathrm{RA},
\end{equation}
where $\hat{\theta} = \diag(i \theta_B, \theta_F)_{\rm BF}$. The transmission distribution function can be obtained from the partition function of the sigma model as \cite{SuppMat}
\begin{gather}
\label{psiBF}
 \psi(\theta_B, \theta_F)
  = \int_{Q(0)}^{Q(L)} DQ e^{-S[Q]}, \\
 \rho(\lambda) = -\frac{2}{\pi} \mathop{\mathrm{Re}} \frac{\partial}{\partial \theta_F} \psi(\theta_B, \theta_F) \Bigr|_{i\theta_B = \theta_F = \pi + 2i\lambda - 0}.
 \label{rhopsi}
\end{gather}
The function $\rho(\lambda)$ is the probability density for the Lyapunov exponent $\lambda$ related to the transmission probability by $T = \cosh^{-2}\lambda$ \cite{Nazarov94, Beenakker97, Khalaf16}. It can be used to compute any moment of the full counting statistics. In this work, we are particularly interested in the average conductance per Weyl node (measured in units of $e^2/h$)
\begin{equation}
 G
  = \int_0^\infty \frac{d\lambda\, \rho(\lambda)}{\cosh^2\lambda}
  = -2 \frac{\partial^2}{\partial \theta_F^2} \psi(\theta_B,\theta_F) \Bigl|_{\theta_B = \theta_F = 0},
 \label{Gpsi}
\end{equation}

\emph{Transfer matrix Hamiltonian.---}
The path integral (\ref{psiBF}) can be computed analytically by mapping to the equivalent Schr\"{o}dinger equation with the position $x$ playing the role of fictitious imaginary time \cite{Feynman65, Efetov83, Efetov99}
\begin{equation}
 \label{Sch}
 \xi\, \frac{\partial \psi}{\partial x}
  = -\mathcal{H} \psi.
\end{equation}
The action (\ref{Snlsm}) describes the motion of a particle on the curved supermanifold parametrized by the matrix $Q$. The presence of chiral modes (second term of the action) results in an effective uniform magnetic field across the manifold with the vector potential $A = -(m/2)\str(T^{-1} \Lambda dT)$. As a result, the transfer matrix Hamiltonian is given by the Laplace-Beltrami operator on the supermanifold with the long derivatives $\partial \mapsto \partial + A$ \cite{SuppMat}. 

The initial conditions (at $x = 0$) force the wave function to be unity at $Q = \Lambda$ and zero everywhere else, see Eq.\ (\ref{BC}). Such a function is invariant under rotating $Q \mapsto K^{-1} Q K$ by any matrix $K$ that commutes with $\Lambda$. This symmetry is preserved by the Hamiltonian $\mathcal{H}$ provided that we compensate the rotation by a gauge transformation: $T \mapsto K^{-1} T K$. As a result, the Hamiltonian conserves the angular momentum corresponding to $K$ rotations and can be restricted to the zero angular momentum sector: $\psi(K^{-1} Q K) = \psi(Q)$ at every $x$. Such functions are parametrized by just two polar angles $0 \leq \theta_F \leq \pi$ and $\theta_B \geq 0$ as in Eq.\ (\ref{BC}). The details of the parametrization and the gauge choice can be found in Supplemental Material \cite{SuppMat}. The explicit effective Hamiltonian has the form
\begin{equation}
 \label{H0}
 \begin{gathered}
  \mathcal{H}
   = -\frac{1}{J} \frac{\partial}{\partial \theta_F} J \frac{\partial}{\partial \theta_F}
     -\frac{1}{J} \frac{\partial}{\partial \theta_B} J \frac{\partial}{\partial \theta_B}
     +\frac{m^2}{4} V(\theta_B,\theta_F), \\
  J
   = \frac{\sin \theta_F \sinh \theta_B}{(\cosh \theta_B - \cos \theta_F)^2},
  \quad
  V
   = \cos^{-2} \frac{\theta_F}{2} - \cosh^{-2} \frac{\theta_B}{2}.
 \end{gathered}
\end{equation}

\emph{Eigenfunctions.---}
The Hamiltonian (\ref{H0}) can be diagonalized with the help of the Sutherland transformation $\tilde{\mathcal{H}} = J^{1/2} \mathcal{H} J^{-1/2} = \tilde{\mathcal{H}}_F + \tilde{\mathcal{H}}_B$ which decouples the variables $\theta_B$ and $\theta_F$ \cite{Rejaei96, Sutherland72, SuppMat}.
The eigenvalues of $\tilde{\mathcal{H}}_F$ form a discrete spectrum with the lowest eigenvalue $(m + 1)^2/4$. The spectrum of $\tilde{\mathcal{H}}_B$ is continuous above zero supplemented by a discrete set of negative eigenvalues (unless $m = 0$) with the lowest at $-(m - 1)^2/4$. 
The overall lowest eigenvalue of $\tilde{\mathcal{H}}$ is $1/4$ in the case $m = 0$, otherwise it equals $m$. The analysis of the full spectrum with the explicit form of the normalized eigenfunctions can be found in Supplemental Material \cite{SuppMat}.

Once the eigenfunctions $\phi_{\nu}$ of the transfer matrix Hamiltonian and the corresponding eigenvalues $\epsilon_\nu$ are known, the solution to the time-evolution problem (\ref{Sch}) can be written explicitly as a spectral expansion \cite{Zirnbauer92, Mirlin94, Rejaei96}
\begin{equation}
 \label{SE}
 \psi(\theta_B, \theta_F, x)
  = \frac{\cos^m(\theta_F/2)}{\cosh^m(\theta_B/2)} + \sum_{\nu} \phi_{\nu} (\theta_B, \theta_F) e^{-(x/\xi) \epsilon_{\nu}}.
\end{equation}
Here the sum over $\nu$ implies a sum over discrete levels and an integral over the continuous part of the spectrum. The eigenfunctions are normalized such that the initial condition is satisfied.

The first term in Eq.\ (\ref{SE}) is the ground state of the transfer matrix Hamiltonian $\mathcal{H}$ with zero eigenvalue. Its contribution to the function $\psi$ and hence to all transport quantities is independent of the length $x$. This zero eigenfunction encodes the effect of chiral topologically protected channels and contributes $m e^2/2h$ to the average conductance. The rest of the spectrum is separated by a finite gap from the ground state and describes the effect of the remaining non-chiral modes. Exponential decay of the second term in Eq.\ (\ref{SE}) signifies localization of the unprotected channels with the gap setting the value of the corresponding localization length $\sim \xi/m$.

\begin{figure}
\center
\includegraphics[width=0.45\textwidth]{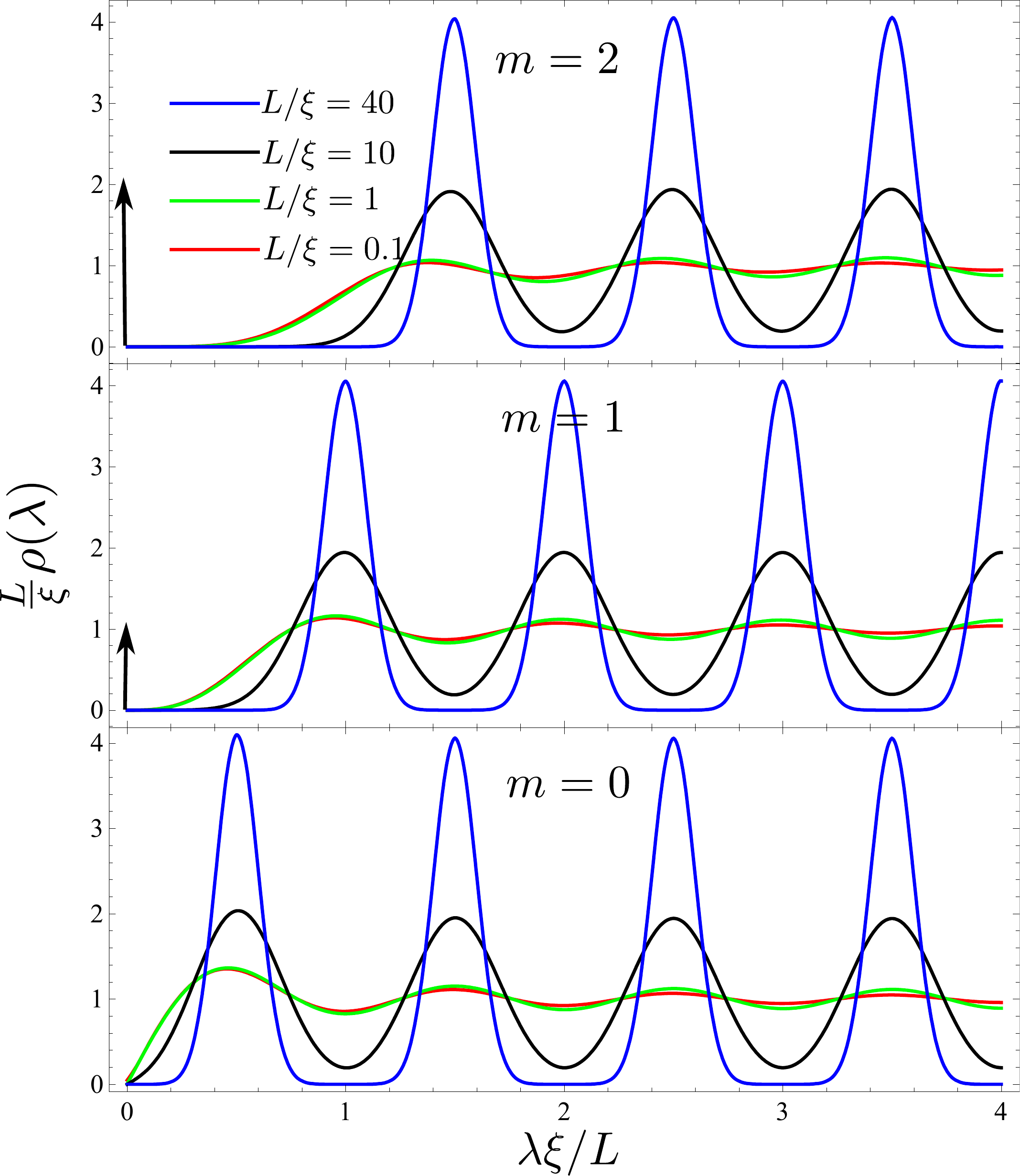}
\caption{Transmission distribution function $\rho(\lambda)$ for several values of $m$ and $L/\xi$. Presence of $m$ chiral channels depletes the distribution in the vicinity of $\lambda = 0$. Level ``crystallization'' occurs at $L \gg \xi$.}
\label{rf}
\end{figure}

\emph{Results.---}
The distribution function $\rho(\lambda)$ is computed from $\psi$ at $x = L$ using Eq.\ (\ref{rhopsi}). The result has the form

\begin{gather}
\rho(\lambda)
 = m\, \delta(\lambda) + \frac{\sinh 2\lambda}{\pi} \!\!\! \sum_{l \in 2\mathbb{N} + 1 + m} \int\limits_{-\infty}^{\infty} dr\; \frac{l  r \; e^{-\frac{L}{4\xi}(l^2 + r^2)}}{l^2 + r^2} \notag\\
 \times  \sinh \frac{\pi}{2}(r-i m)  \; P^{(-m,0)}_{\frac{ir+m-1}{2}}(\cosh 2\lambda) P^{(0,m)}_{\frac{l-m-1}{2}}(-\cosh 2\lambda).
 \label{rho}
\end{gather}
Here $\mathbb{N} = 0,1,\dots$ and $P^{(\alpha,\beta)}_\nu(x)$ is the Jacobi polynomial of (possibly complex) order $\nu$ \cite{SuppMat}. The term $m \delta(\lambda)$ represents the contribution of $m$ chiral channels with ideal transparency.

The distribution function (\ref{rho}) is plotted in Fig.\ \ref{rf} for several values of $m$ and $L/\xi$. Qualitatively, we observe (i) a suppression of the distribution close to $\lambda = 0$ (which corresponds to ideal transmission) and (ii) ``crystallization'' of individual transmission eigenvalues in the limit $L \gg \xi$ \cite{Frahm95, Lamacraft04}. Both effects can be interpreted in terms of statistical level repulsion. Different Lyapunov exponents $\lambda$ repel each other with strength proportional to $L/\xi$. The presence of $m$ eigenvalues pinned at $\lambda = 0$ pushes the remaining spectrum away from zero.

In the limit $L \gg \xi$, the distribution function is a sum of equidistant Gaussian peaks with width $\sqrt{L/\xi}$ separated by $L/\xi$ and having a unit weight. The position of the first peak determines the localization length governing the typical conductance (of non-chiral modes) $G_{\text{typ}} = e^{\left<\ln G \right>} \propto e^{- \frac{L}{\xi}(m+1)}$ leading to $\xi^{\text{typ}} = \xi/(1 + m)$. 


\begin{figure}
\center
\includegraphics[width=0.46 \textwidth]{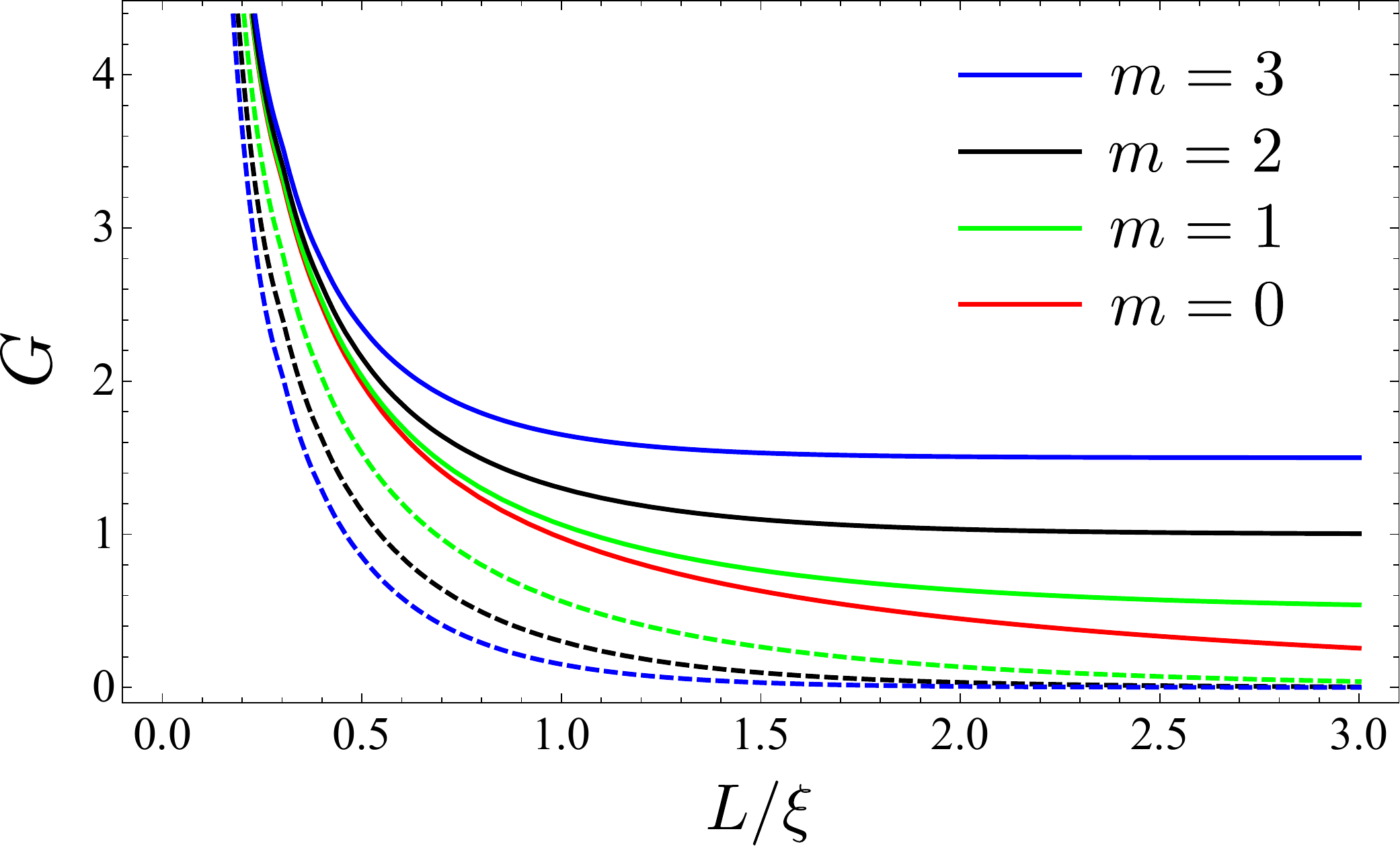}
\caption{Average conductance for different values of $m$ as a function of $L/\xi$ (solid) and the contribution of the non-chiral modes $G - G_{\infty}$ (dashed). For $L \ll \xi$, the behavior is metallic $G \sim \xi/L$, receiving contribution from all channels. For $L \gg \xi$, the conductance is solely due to the chiral channels and approaches $me^2/2h$.}
\label{Gf}
\end{figure}

The average conductance per Weyl node is given by Eq.\ (\ref{Gpsi}) (in units of $e^2/h$)
\begin{equation}
\label{G}
G = \frac{m}{2} + \! \! \! \!\sum_{l\in 2\mathbb{N}+1+m} \int \limits_{-\infty-im}^{\infty-im} \!\!\!\! dr \frac{l r \tanh \frac{\pi}{2}(r+i m)}{l^2 + r^2} e^{-\frac{L}{4\xi}(l^2 + r^2)}.
\end{equation}
It is shown in Fig.\ \ref{Gf} as a function of $L/\xi$ for different values of $m$. A smooth crossover from the diffusive limit $L \ll \xi$ to the strongly localized regime $L \gg \xi$ where current is carried only by the protected chiral channels is clearly visible. In the latter case, the conductance is given by
\begin{equation}
\label{G_LW}
 G
  = \frac{m}{2} + \begin{cases}
      2 (\pi \xi/L)^{3/2}\,  e^{-L/4\xi}, & m = 0, \\
      2 \sqrt{\xi/\pi L}\,  e^{-L/\xi}, & m = 1, \\
      \dfrac{m^2 - 1}{m}\, e^{-mL/\xi}, & m \geq 2.
    \end{cases}
\end{equation}
The localization length governing the decay of the average conductance (for the non-chiral modes) is generally different from the typical localization length \cite{Evers08}. For $m \neq 0$, it is significantly reduced: $\xi^{\text{av}} = \xi/4m$.

\begin{figure}
\center
\includegraphics[width=0.45 \textwidth]{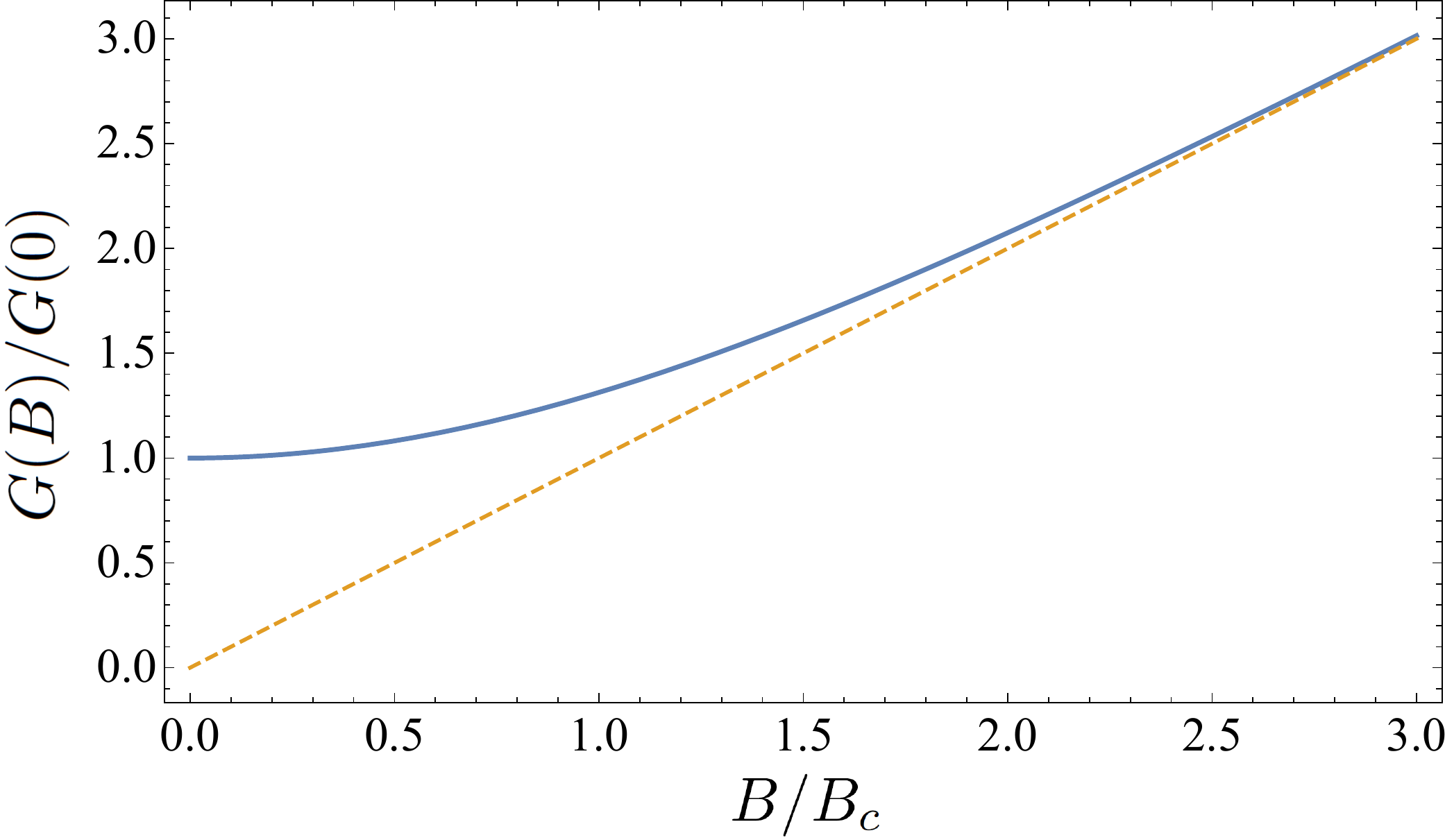}
\caption{Average magnetoconductance in the limit $L \ll \xi$. A crossover from quadratic to linear field dependence occurs at $B_c = 2\Phi_0 \xi/L A = 2\sigma \Phi_0/L$.}
\label{GBf}
\end{figure}

Let us now discuss the longitudinal magnetoconductance of a WSM sample with length $L \ll \xi = \sigma A$. Magnetic field is directly related to the parameter $m = B A/\Phi_0$ (here $B$ is the component of magnetic field in the direction of current). A crossover from quadratic to linear magnetoconductance occurs when $B$ exceeds $B_c = 2\Phi_0 \xi/L A = 2\sigma \Phi_0/L$. It corresponds to $m = \xi/L \gg 1$ hence discreteness of $m$ can be disregarded. The field $B_c$ decreases with increasing disorder strength. In this limit, the magnetoconductance (Fig.\ \ref{GBf}) reduces to the semiclassical result \cite{Khalaf16, Altland16}:
\begin{equation}
\label{dG}
 \frac{G(B)}{G(0)}
  = \frac{B}{B_c} \coth \frac{B}{B_c}.
\end{equation}

\emph{Discussion.---}
In the presence of chiral channels, localization effects in the non-chiral bands are so strong that they can be observed in a 3D WSM sample. Both the length $\xi = \sigma A$ and the Landau level degeneracy $m = B A/\Phi_0$ scale linear with $A$. Hence the typical localization length $\xi^\text{typ} = \sigma \Phi_0/B$ for the non-chiral modes remains finite for arbitrary sample cross-section.

In a field stronger than $B_c = 2\sigma \Phi_0/L$, all the non-chiral modes are effectively localized and the current is carried only by the chiral channels representing the lowest Landau level. The value of $B_c$ is considerably smaller than the ultra-quantum field $B_u = \left(\mu/\hbar v \right)^2 \Phi_0/4\pi$ at which only the lowest Landau level lies below the Fermi energy $\mu$. Both $B_c$ and $B_u$ scale as $\mu^2$ and their ratio is $B_c/B_u \sim l/L \ll 1$. This implies that the ultra-quantum limit for transport in a disordered WSM occurs at much weaker fields than in the ideal clean material. The main experimental manifestation of this effect would be observing a linear longitudinal magnetoconductance coexisting with quantum oscillations indicating multiple Landau levels at the Fermi surface.

Our results are valid as long as scattering between different Weyl nodes is neglected. This is the case when the dominant scatterers are smooth on the scale of the lattice spacing and the Weyl nodes are well-separated in momentum space. Coupling between the Weyl nodes (e.g. by sharp impurities) introduces a length scale $l_i$. The effect of introducing this length scale would be replacing the sample length $L$ with $l_i$ in our results for magnetoconductivity. In particular, Eq. (\ref{dG}) reproduces the quadratic $B$ dependence obtained from the kinetic equation treatment \cite{Son13} in the limit of small $B$. The transport ultra-quantum field is in this case related to the clean ultra-quantum field by $B_c/B_u \sim l/l_i$ which is small for sufficiently smooth disorder \cite{footnote2}.


\emph{Conclusion.---}
We have derived exact expressions for longitudinal conductance, Eq.\ (\ref{G}), and transmission distribution function, Eq.\ (\ref{rho}), in a Weyl semimetal subject to external magnetic field. The presence of the chiral topologically protected Landau level with the macroscopic degeneracy $m = BA/\Phi_0$ leads to a very efficient localization of the non-chiral states. An effective ultra-quantum regime, where only the chiral modes contribute to transport, occurs at a length scale $\xi^\text{typ} \approx \xi/m = \sigma\Phi_0/B$, see Fig.\ \ref{Gf}. Magnetoconductance changes from quadratic to linear at the disorder-dependent scale $B_c = 2\sigma \Phi_0/L$, Fig.\ \ref{GBf}, that is significantly smaller than the ultra-quantum field for a clean sample, $B_c/B_u \sim l/L \ll 1$. Thus magnetic field dramatically enhances disorder effects.

The same results apply in general to any quasi-1D system with broken time-reversal symmetry with a number of chiral modes $m$. The transmission distribution function, Fig.\ \ref{rf}, interpolates between diffusive ($L \ll \xi$) and strongly localized ($L \gg \xi$) limits exhibiting ``crystallization'' of the Lyapunov exponents in the latter case. This effect can be interpreted in terms of statistical level repulsion. 

\begin{acknowledgments}
\emph{Acknowledgments.---}
We are grateful to D.\ Bagrets, A.\ Mirlin, M.\ Skvortsov, and A.\ Stern for valuable and stimulating discussions. The work was supported by the Russian Science Foundation (Grant No. 14-42-00044).
\end{acknowledgments}


\supplementarystart

\centerline{\bfseries\large ONLINE SUPPLEMENTAL MATERIAL}
\vspace{6pt}
\centerline{\bfseries\large Localization effects on magnetotransport of a Weyl semimetal}
\vspace{6pt}
\centerline{E.~Khalaf and P.\,M.~Ostrovsky}
\begin{quote}
In this Supplemental Material, we provide technical details that are relevant for the text of the Letter. First, we present a derivation of the non-linear sigma model for the matrix Green's function in the presence of chiral channels. We then provide a detailed derivation of the transfer matrix Hamiltonian corresponding to the non-linear sigma model with topological term, Eq.\ (2). We consider an explicit parametrization of the sigma-model manifold and show that the transfer matrix Hamiltonian reduces to Eq.\ (8) for a specific gauge choice. Finally, we give explicit expressions for the eigenfunctions of the transfer matrix Hamiltonian (8) and the spectral expansion of the evolution operator $\psi(\theta_B, \theta_F,x)$.
\end{quote}


\section{Sigma model Derivation}
In this section, we explain briefly how the non-linear sigma model for the matrix Green's function of a quasi-1D wire with both chiral and non-chiral channels is derived. For a more detailed derivation, the reader is referred to Ref. \cite{SKhalaf16}.

The transport properties of a quasi-1D wire placed between two perfect metallic leads are fully determined by the matrix of transmission amplitudes $t_{mn}$ which acts in the space of 1D conducting channels. The main observables of interest in this work are the conductance and the distribution function of transmission eigenvalues $\rho(\lambda)$. The latter gives the probability to find a channel with transmission probability $\cosh^{-2}\lambda$. Both quantities can be written in terms of the transmission matrix $t_{mn}$ as
\begin{equation}
 G
  = \tr t^{\dagger} t,
 \qquad
 \rho(\lambda)
  = 2\frac{\tanh\lambda}{\cosh^2\lambda} \tr \delta\bigl( \cosh^{-2}\lambda - t^\dagger t \bigr).
\end{equation}
Here, as in the main text, conductance is measured in units of $e^2/h$ per Weyl node.

The distribution function of transmission eigenvalues can be obtained from the matrix Green's function defined as \cite{SNazarov94}
\begin{equation}
\label{MGFF_GN}
 \begin{pmatrix}
   \epsilon - \hat{H} + i 0 & \hat{v} \sin(\theta/2) \, \delta(x - x_L) \\
   \hat{v} \sin(\theta/2) \, \delta(x - x_R) & \epsilon - \hat{H} - i 0
 \end{pmatrix} \check G(x,x',\theta)
  = \delta(x - x') \check{\mathbbm{1}}.
\end{equation}
where $x_L$ and $x_R$ are points in the left and right leads respectively and $\check{\mathbbm{1}}$ in the right-hand side is unity in the channel and matrix retarded-advanced (RA) spaces. The introduction of the counting field $\theta$ makes it possible to obtain the full transmission distribution function using a single Green's function. 

The Hamiltonian $\hat{H}$ acts in both 1D real space ($x$ variable) and channel space and it is given in terms of the velocity operator $\hat v$ (which acts in the channel space only) by
 \begin{equation}
 \label{NLSM_H}
 \hat{H} = - i \hat{v}  \otimes \frac{\partial}{\partial x} + \hat V(x), \qquad \hat{v} = \begin{pmatrix} \mathbbm{1}_{n_R} & 0 \\ 0 & -\mathbbm{1}_{n_L} \end{pmatrix}.
 \end{equation}
Here, $\hat V(x)$ is the disorder potential. It is assumed to be a random Hermitian matrix obeying Gaussian
 distribution with $\langle \hat V \rangle = 0$ and $\langle V_{nm}(x) V_{mn}(x') \rangle = \frac{1}{N \tau}\, \delta(x - x')$, where $\tau$ is the electron mean free time. 

Transport properties of a disordered system are calculated by averaging over disorder realizations. This is most easily achieved with the help of the supersymmetric representation. Let us consider two matrix Green functions (\ref{MGFF_GN}) with different source parameters $i \theta_B$ and $\theta_F$ and define the partition function as
\begin{equation}
 \label{ZBF}
 \psi(\theta_B, \theta_F)
  = \frac{\det \check G(i \theta_B)}{\det \check G(\theta_F)}
  = \frac{\det\bigl[ 1 - \sin^2(\theta_F/2) t^{\dagger} t \bigr]}{\det\bigl[ 1 + \sinh^2(\theta_B/2) t^{\dagger} t \bigr]}.
\end{equation}
This quantity can be viewed as a superdeterminant if the Green function $\check G$ is extended into a superspace. The two parameters $\theta_{B,F}$ have the meaning of bosonic and fermionic counting fields. The distribution function $\rho(\lambda)$ and the conductance $G$ can be expressed in terms of $\psi$ as
\begin{equation}
 \rho(\lambda) = -\frac{2}{\pi} \mathop{\mathrm{Re}} \frac{\partial}{\partial \theta_F} \psi(\theta_B, \theta_F) \Bigr|_{i\theta_B = \theta_F = \pi + 2i\lambda - 0}, \qquad G = -2 \frac{\partial^2}{\partial \theta_F^2} \psi(\theta_B,\theta_F) \Bigl|_{\theta_B = \theta_F = 0}.
\end{equation}

We consider a system with $N$ channels in total (both left- and right-moving) and $m$ chiral channels (moving in one direction only). To perform the disorder average, we write the partition function $\psi(\theta_B, \theta_F)$ as a Gaussian integral over superfields,
\begin{gather}
 \label{NLSM_ZFB}
 \psi(\theta_B, \theta_F) = \int D\phi^{\dagger}\, D\phi\; e^{-S[\phi^\dagger, \phi]}, \\
 S = -i \int dx\, \phi^{\dagger} \Lambda  \left(\epsilon - \hat{H} + i 0 \Lambda + M \right) \phi. \label{Spsi}
\end{gather}
Here the supervector $\phi$ contains $2N$ complex and $2N$ Grassmann variables. It operates in retarded-advanced (RA), Bose-Fermi (BF), and channel spaces. The matrix $\Lambda$ is defined in (2), whereas $M$ represents the source terms in Eq.\ (\ref{MGFF_GN}). The Gaussian integral in Eq.\ (\ref{NLSM_ZFB}) yields the superdeterminant of the corresponding matrix, which has the form of the ratio of usual determinants from Eq.\ (\ref{ZBF}).

The next step is the averaging over disorder yielding the quartic term $(\phi^\dagger \phi)^2$. This term is decoupled via a Hubbard-Stratonovich transformation introducing the supermatrix field $Q$ that acts in RA and BF spaces but not in the channel space. The action is then integrated over the supervector $\phi$ leading to
\begin{equation}
\label{NLSM_trln}
 S[Q]
  = \frac{N}{8 \tau} \Str Q^2
    + \frac{N-m}{2} \Str \ln \left(i v\, \frac{\partial}{\partial x} + \frac{i Q}{2\tau} \right)
    + \frac{N+m}{2} \Str \ln \left(-i v\, \frac{\partial}{\partial x} + \frac{i Q}{2\tau} \right).
\end{equation}
Here, ``Str'' denotes the operator supertrace over BF and RA spaces including integration in the real space and $\tau$ is the mean free time.

For large number of channels $N \gg 1$, we can treat the action (\ref{NLSM_trln}) within the saddle-point approximation \cite{SEfetov99}. Assuming the matrix $Q$ is constant in space, we identify a degenerate minimum of the action $Q = T^{-1} \Lambda T$ with any supermatrix $T$ acting in BF and RA spaces. The saddle point manifold is determined by the requirement that the $Q$ integral is convergent \cite{SMirlin00}. This restricts the matrix $T$ to the compact group manifold $\mathrm{U}(2)$ in the fermionic sector and non-compact $\mathrm{U}(1,1)$ group in the bosonic sector. Hence, $T$ belongs to the superunitary group $\mathrm{U}(1,1|2)$. The matrix $Q$, parametrized as $T^{-1} \Lambda T$, is invariant under multiplying $T$ from the left by any matrix $K$ that commutes with $\Lambda$. These matrices from a subgroup of the superunitary group $\mathrm{U}(1,1|2)$ that can be identified with $\mathrm{U}(1|1) \times \mathrm{U}(1|1)$. As a result, the matrix $Q$ belongs to the coset space $\mathrm{U}(1,1|2) / \mathrm{U}(1|1) \times \mathrm{U}(1|1)$. Its compact (FF) and non-compact (BB) parts have the form of a sphere $\mathrm{U}(2)/ \mathrm{U}(1) \times \mathrm{U}(1) \simeq S^2$ and a hyperboloid $\mathrm{U}(1,1)/ \mathrm{U}(1) \times \mathrm{U}(1) \simeq H^2$, respectively. 

The effective low-energy theory is derived by a gradient expansion of Eq.\ (\ref{NLSM_trln}) assuming that $T(x)$ varies slowly in space. With a cyclic permutation of matrices under the supertrace, the action can be written in the form
\begin{equation}
 \label{gradexp}
 S[Q]
  = \frac{N-m}{2} \Str \ln \left(i v\, \frac{\partial}{\partial x} - i v \frac{\partial T}{\partial x} T^{-1} + \frac{i \Lambda}{2\tau} \right) 
    + \frac{N+m}{2} \Str \ln \left(-i v\, \frac{\partial}{\partial x} + i v \frac{\partial T}{\partial x} T^{-1} + \frac{i \Lambda}{2\tau} \right).
\end{equation}
Expanding the logarithms in Eq.\ (\ref{gradexp}) up to the second order in small derivatives $(\partial T/\partial x) T^{-1}$, we obtain the action of the sigma model given in (2). Appearance of the topological (second) term in Eq.\ (2) is thus directly related to the imbalance $m$ between left- and right-moving channels. This term is exactly what appears in the field theory of a quantum Hall edge by partial integration of the Pruisken $\theta$-term \cite{SPruisken84}. The fact that it is written in terms of $T$ rather than $Q$ reflects its gauge dependence which follows from the fact that theory (2) describes a single Weyl node. Including a Weyl node with opposite chirality restores the gauge invariance of the theory.

The boundary conditions (3) can be traced back to the off-diagonal terms containing the counting fields in (\ref{MGFF_GN}). They follow from the requirement of continuity of the $Q$ matrix at $x=0$ and $L$. The counting fields $\theta_{B,F}$ enter the sigma model only via these boundary conditions \cite{SKhalaf16}. The partition function (\ref{ZBF}) is given by the path integral in the superspace of the matrix $Q$ with the action (2). All the paths start at the point $Q = \Lambda$, representing the ``north pole'' of the sphere in the compact FF sector and the base of the hyperboloid in the non-compact BB sector, and end in a point with the polar angles $\theta_F$ and $\theta_B$ on the sphere and the hyperboloid, respectively, determined by the counting fields.

\section{Transfer matrix method}

In this section, we introduce the transfer matrix method which is used to solve the path integral (2) exactly. The transfer matrix method is based on the observation that a 1D field theory can be reduced to a time-evolution Schr\"{o}dinger problem \cite{SFeynman65} with the position $x$ playing the role of time. We find it convenient to define the time variable $t$ as the length along the wire measured in units of $\xi$. The derivation is a generalization of the standard derivation of the Sch\"{o}dinger equation from the path integral in quantum mechanics \cite{SFeynman65}.

We start with the action (2) and choose some coordinates on the sigma-model supermanifold (explicit parametrization is given in the next subsection) which is the coset space $\U(1,1|2)/\U(1|1) \times \U(1|1)$. It is parametrized by 4 real and 4 Grassmann coordinates that we denote by $y^{\alpha}$ with $\alpha = 1,\ldots,8$. Commutation relations have the form $y^{\alpha} y^{\beta} = s_{\alpha \beta} y^{\beta} y^{\alpha}$ with $s_{\alpha \beta}$ defined as
\begin{equation}
\label{3T_sab}
 s_{\alpha \beta}
  = \begin{cases}
      -1, & \text{both $\alpha$ and $\beta$ refer to Grassmann coordinates}, \\
      1, & \text{otherwise}.
    \end{cases}
\end{equation}
In the following, we will assume that repeated indices (one upper and one lower) are summed over, but that does not include the indices of $s_{\alpha \beta}$.

In terms of the coordinates $y^{\alpha}$, the action (2) has the form
\begin{equation}
\label{3T_SgA}
 S
  = \int_0^{L/\xi} dt \left(
      \frac{1}{4} \dot{y}^{\alpha} \dot{y}^{\beta} g_{\beta \alpha} + \dot{y}^{\alpha} A_{\alpha}
    \right),
\end{equation}
where we used the ``dot'' to denote a derivative with respect to the time variable $t$. By comparison to (2), it is easy to see that the metric tensor and the vector potential are given by the relations
\begin{gather}
\label{3T_g}
 dy^{\alpha}\, dy^{\beta} g_{\beta \alpha} = - \frac{1}{2} \str (dQ)^2, \\
\label{3T_A}
 dy^{\alpha} A_{\alpha} = -\frac{m}{2} \str T^{-1} \Lambda  dT.
\end{gather}
The symmetry properties of the metric tensor follow from (\ref{3T_g}): it satisfies the identity $g_{\alpha \beta} = s_{\alpha \beta} g_{\beta \alpha}$.

\subsection{Derivation of the transfer matrix Hamiltonian}

The action (\ref{3T_SgA}) describes the motion of a particle in a curved supermanifold with metric tensor $g_{\alpha \beta}$ in the presence of vector potential $A_{\alpha}$. As a result, we expect the transfer matrix Hamiltonian operator which generates the time evolution governed by (\ref{3T_SgA}) to be given by the Laplace-Beltrami operator on the supermanifold with long derivatives $\partial_{\alpha} \rightarrow \partial_{\alpha} + A_{\alpha}$. This is what we are going to verify below. The main subtlety in the following derivation lies in the extra minus signs which appear while handling anticommuting variables.

The derivation of the transfer matrix Hamiltonian, corresponding to the action (\ref{3T_SgA}), starts with writing the imaginary time evolution equation
\begin{equation}
 \psi(y,t)
  = \int dy' \int_{(y',t')}^{(y,t)} D y  \;  e^{-S[y(t)]}\psi(y',t').
 \label{stepint}
\end{equation}
We consider the evolution for infinitesimal time interval $t' = t - \epsilon$ and restrict the path integral between $y$ and $y'$ to the classical trajectory. Such a trajectory obeys the Euler-Lagrange equation. Minimizing the action (\ref{3T_SgA}), we have
\begin{equation}
 \ddot{y}^{\beta} g_{\beta \alpha}
  = \dot{y}^{\beta} \dot{y}^{\gamma} \left(
      \frac{s_{\alpha\beta} s_{\alpha\gamma}}{2}\, \partial_\alpha g_{\gamma \beta} - \partial_\gamma g_{\beta \alpha}
    \right)
    + 2 \dot{y}^{\beta} \Bigl( s_{\alpha\beta} \partial_\alpha A_{\beta} - \partial_\beta A_{\alpha} \Bigr).
 \label{EL}
\end{equation}
We expand the action of the infinitesimal trajectory to the second order in small $\epsilon$ and express the result in terms of the coordinates and velocities at the time $t$:
\begin{align}
 S
  &= \int_{t-\epsilon}^t dt\, \left(
      \frac{1}{4}\, \dot{y}^{\alpha} \dot{y}^{\beta} g_{\beta \alpha} + \dot{y}^{\alpha} A_{\alpha}
    \right) \nonumber \\
  &= \frac{\epsilon}{4}\, \dot{y}^{\alpha} \dot{y}^{\beta} g_{\beta \alpha}
    +\epsilon \dot{y}^{\alpha} A_{\alpha}
    -\frac{\epsilon^2}{8} \Bigl[
      \dot{y}^{\alpha}  \dot{y}^{\beta} \dot{y}^{\gamma} \partial_{\gamma} g_{\beta \alpha}
      +2\dot{y}^{\alpha} \ddot{y}^{\beta} g_{\beta \alpha}
    \Bigr] -\frac{\epsilon^2}{2} \Bigl(
      \ddot{y}^{\alpha} A_{\alpha}
      + \dot{y}^{\alpha} \dot{y}^{\beta} \partial_{\beta} A_{\alpha}
    \Bigr) \nonumber \\
  &= \frac{\epsilon}{4}\, \dot{y}^{\alpha} \dot{y}^{\beta} g_{\beta \alpha}
    +\epsilon \dot{y}^{\alpha} A_{\alpha}
    -\frac{\epsilon^2}{2} \Bigl(
      \ddot{y}^{\alpha} A_{\alpha}
      + \dot{y}^{\alpha} \dot{y}^{\beta} \partial_{\beta} A_{\alpha}
    \Bigr).
\label{Sexpand}
\end{align}
Here $\partial_\alpha = \partial/\partial y^\alpha$. Substituting $\ddot y$ from Eq.\ (\ref{EL}), we see that the term in square brackets vanishes identically hence we have dropped it in the final expression.

Let us now analyze Eq.\ (\ref{stepint}) in the leading order $O(\epsilon^0)$. We parametrize the trajectory by the velocity $\dot y$ at time $t$ rather than by the point $y'$. In the leading order, we retain only the first term in the expansion (\ref{Sexpand}) and replace $\psi(y', t')$ by $\psi(y, t)$. The equation is fulfilled provided the integration measure is $dy' = \sqrt{|g|} d \dot y$, such that the Gaussian integral over $\dot y$ yields unity (here $|g|$ is the superdeterminant of $g$).
\begin{equation}
 \int \sqrt{|g|}\, d\dot y\, \exp\left(-\frac{\epsilon}{4}\, \dot{y}^{\alpha} \dot{y}^{\beta} g_{\beta \alpha}\right)
  = 1.
 \label{Gauss}
\end{equation}
Thus we have established the proper integration measure in Eq.\ (\ref{stepint}) in terms of $\dot y$. 

Typical values of the velocity in the Gaussian integral (\ref{Gauss}) are $\dot y \sim \epsilon^{-1/2}$. From Eq.\ (\ref{EL}) we conclude that $\ddot y \sim (\dot y)^2 \sim \epsilon^{-1}$. This allows us to classify the terms of the expansion (\ref{Sexpand}) in powers of $\epsilon$. The first (leading) term is effectively $O(\epsilon^0)$, the second term is $O(\epsilon^{1/2})$, and the last term is $O(\epsilon)$. Thus the expansion (\ref{Sexpand}) is accurate up to linear order in $\epsilon$.

Now we consider Eq.\ (\ref{stepint}) up to the terms $O(\epsilon)$. In order to do this, we expand the wave function $\psi(y', t - \epsilon)$ as
\begin{equation}
 \psi(y', t')
  = \psi\left(y - \epsilon \dot{y} + \frac{\epsilon^2}{2} \ddot{y}, t-\epsilon\right) 
  = \left[
      1 - \epsilon \left( \frac{\partial}{\partial t} + \dot{y}^{\alpha} \partial_{\alpha} \right)
      +\frac{\epsilon^2}{2} \bigl( \ddot{y}^{\alpha} \partial_{\alpha} + \dot{y}^{\alpha} \dot{y}^{\beta} \partial_{\beta} \partial_{\alpha} \bigr)
    \right] \psi(y, t).
 \label{step}
\end{equation}
Applying the same estimates as above, we see that this expansion is effective up to $O(\epsilon)$. Substituting the expansions (\ref{step}) and (\ref{Sexpand}) into Eq.\ (\ref{stepint}) and equating the terms of the order $O(\epsilon)$, we obtain the evolution equation for the function $\psi$ in the form of an imaginary time Schr\"{o}dinger equation [cf.\ Eq.\ (7)] 
\begin{equation}
 \frac{\partial\psi}{\partial t}
  = -\mathcal{H} \psi,
 \qquad
 \mathcal{H}
  = -\frac{\epsilon}{2} \Bigl[
      \langle \dot{y}^{\alpha} \dot{y}^{\beta} \rangle (\partial_{\beta} + A_{\beta})(\partial_{\alpha} + A_{\alpha})
      +\langle \ddot{y}^{\alpha}\rangle (\partial_{\alpha} + A_{\alpha})
    \Bigr].
 \label{3T_Hi}
\end{equation}
Here $\langle\ldots\rangle$ denotes a Gaussian average with the weight (\ref{Gauss}).

The average $\langle \dot{y}^{\alpha} \dot{y}^{\beta}\rangle$ can be easily evaluated as 
\begin{equation}
\label{3T_dydy}
 \langle \dot{y}^{\alpha} \dot{y}^{\beta}\rangle
  = \frac{2}{\epsilon} s_{\alpha \alpha} g^{\alpha \beta},
\end{equation}
where $g^{\alpha \beta}$ is the inverse metric tensor: $g_{\alpha \beta} g^{\beta \gamma} = \delta_{\alpha}^{\gamma}$. Note that this inverse tensor has different symmetry properties, $g^{\alpha \beta} = s_{\alpha \alpha} s_{\beta \beta} s_{\alpha \beta} g^{\beta \alpha}$. To compute $\langle\ddot{y}^{\alpha}\rangle$, we take only the first term from the equation of motion (\ref{EL}) since the second term is effectively of the higher order $O(\epsilon^{-1/2})$. This yields
\begin{align}
\label{3T_ddy}
 \langle \ddot{y}^{\alpha} \rangle
  &= \langle \dot{y}^{\gamma} \dot{y}^{\mu} \rangle \left(
      \frac{s_{\beta\gamma} s_{\beta \mu}}{2}\, \partial_\beta g_{\mu \gamma} - \partial_\mu g_{\gamma \beta}
    \right) g^{\beta \alpha}
  = \frac{1}{\epsilon} \Bigl(
      s_{\beta\gamma} s_{\beta \mu} s_{\gamma \gamma} g^{\gamma \mu} \partial_\beta g_{\mu \gamma} - 2 s_{\gamma \gamma} g^{\gamma \mu} \partial_\mu g_{\gamma \beta}
    \Bigr) g^{\beta \alpha} \nonumber \\
  &= \frac{1}{\epsilon} \Bigl(
      \partial_\beta \ln |g| + 2 \partial_\mu g^{\mu \alpha}
    \Bigr)
  = \frac{2}{\epsilon \sqrt{|g|}} \partial_{\beta} \bigl( \sqrt{|g|} g^{\beta \alpha} \bigr).
\end{align}
Substituting (\ref{3T_dydy}) and (\ref{3T_ddy}) in (\ref{3T_Hi}), we finally obtain the Hamiltonian
\begin{equation}
\label{H}
 \mathcal{H}
  = -\frac{1}{\sqrt{|g|}} (\partial_{\alpha} + A_{\alpha}) \sqrt{|g|} g^{\alpha \beta} (\partial_{\beta} + A_{\beta}).
\end{equation}
This is exactly the Laplace-Beltrami operator on the supermanifold \cite{SMirlin94} with long derivatives $\partial_{\alpha} \mapsto \partial_{\alpha} + A_{\alpha}$.

\subsection{Parametrization and gauge choice}

In order to proceed further, we make an explicit choice of coordinates for the coset space $\U(1,1|2)/\U(1|1) \times \U(1|1)$. We define the four real parameters $0 \leq \theta_F \leq \pi$, $\theta_B >0$, $0 \leq \phi_{B,F} < 2\pi$ and the four Grassmann coordinates $\mu$, $\nu$, $\eta$, and $\sigma$. An explicit parametrization of the matrices $T$ and $Q$ for Eq.\ (2) expresses them in terms of the coordinate vector $y^{\alpha} = (\theta_B, \theta_F, \phi_B, \phi_F, \mu, \nu, \eta, \sigma)^T$.

We choose the parametrization for the $T$ matrix as
\begin{equation}
\begin{gathered}
 T
  = K^{-1} \begin{pmatrix} \cos(\hat\theta/2) & i \sin(\hat\theta/2) \\ i \sin(\hat\theta/2) & \cos(\hat\theta/2) \end{pmatrix}_\text{RA} K,
 \qquad
 \hat{\theta}
  = \begin{pmatrix} i \theta_B & 0 \\ 0 & \theta_F \end{pmatrix}_\text{BF},
 \qquad
 K
  = \begin{pmatrix} e^{i \hat{\phi}}\, u & 0 \\ 0 & e^{-i \hat{\phi}}\, v \end{pmatrix}_\text{RA}, \\
 \hat{\phi}
  = \begin{pmatrix} \phi_B & 0 \\ 0 & \phi_F \end{pmatrix}_\text{BF},
 \qquad
 u
  = \exp \begin{pmatrix} 0 & \mu \\ \nu & 0 \end{pmatrix}_\text{BF},
 \qquad
 v
  = \exp \begin{pmatrix} 0 & \eta \\ \sigma & 0 \end{pmatrix}_\text{BF}.
\end{gathered}
\label{Tpar}
\end{equation}
Note that here the gauge is chosen such that the matrix multiplying $T$ from the left is the inverse of $K$ multiplying $T$ from the right. This gauge is singular only at the ``south pole'' $\theta_F = \pi$ and represents a Dirac string going through this point. Parametrization of $Q$ follows from Eq.\ (\ref{Tpar}) through the relation $Q = T^{-1} \Lambda T$. 

With the parametrization (\ref{Tpar}), the metric tensor takes the block-diagonal form
\be
\label{g}
g = 
\left(\begin{array}{cc} 
\mathbbm{1}_2 & 0 \\
0 & g_K
\end{array} \right).
\ee
The $6 \times 6$ block $g_K$ and the $8$-component vector potential $A$ are given explicitly by
\begin{gather}
g_K = 
\left(\begin{array}{cccccc} 
x_B^2 & 0 & -\frac{i}{2} \nu x_B^2 & -\frac{i}{2} \mu x_B^2 & \frac{i}{2} \sigma x_B^2 & \frac{i}{2} \eta x_B^2 \\
0 & x_F^2 & -\frac{i}{2} \nu x_F^2 & -\frac{i}{2} \mu x_F^2 & \frac{i}{2} \sigma x_F^2 & \frac{i}{2} \eta x_F^2 \\
-\frac{i}{2} \nu x_B^2 & -\frac{i}{2} \nu x_F^2 & 0 & -1 + y_B y_F + \frac{r^2}{4} \mu \nu & -\frac{r^2}{4} \sigma \nu &  i e^{i \chi} x_B x_F - \frac{r^2}{4} \eta \nu \\
 -\frac{i}{2} \mu x_B^2 & -\frac{i}{2} \mu x_F^2 &  1 - y_B y_F + \frac{r^2}{4}\nu \mu & 0  & -i e^{-i \chi} x_B x_F - \frac{r^2}{4}\sigma \mu & -\frac{r^2}{4} \eta \mu \\
\frac{i}{2} \sigma x_B^2 & \frac{i}{2} \sigma x_F^2 & -\frac{r^2}{4} \nu \sigma &  i e^{-i \chi} x_B x_F - \frac{r^2}{4} \mu \sigma & 0 & -1 + y_B y_F + \frac{r^2}{4}\eta \sigma \\
\frac{i}{2} \eta x_B^2 & \frac{i}{2} \eta x_F^2  &  -i e^{i \chi} x_B x_F - \frac{r^2}{4} \nu \eta & -\frac{r^2}{4} \mu \eta & 1 - y_B y_F + \frac{r^2}{4} \sigma \eta & 0
\end{array} \right), \nonumber \\
A = \frac{m}{2} \left[ 0,0,i (y_B - 1), i (1 - y_F), \frac{1}{2} \nu (y_B - y_F),\frac{1}{2} \mu (y_B - y_F),-\frac{1}{2} \sigma (y_B - y_F),-\frac{1}{2} \eta (y_B - y_F) \right]^T, \nonumber \\
x_B = \sinh \theta_B, \quad x_F = \sin \theta_F, \quad y_B = \cosh \theta_B, \quad y_F = \cos \theta_F, \quad r=\sqrt{x_B^2 + x_F^2}, \quad \chi = \phi_F - \phi_B.
\label{gKA}
\end{gather}
The main observation here is that the metric tensor decouples into a part in the polar angles $(\theta_B, \theta_F)$, that is just the identity matrix, and a part in the $K$-variables $(\phi_B, \phi_F, \mu, \nu, \eta, \sigma)$, that has the complicated form (\ref{gKA}), while the radial components of the vector potential vanish: $A_{\theta_B} = A_{\theta_F}  = 0$.

The Jacobian is computed as the square root of the superdeterminant of the metric tensor giving (up to a constant factor)
\be
\label{J}
J = \sqrt{|g|} = \frac{\sin \theta_F \sinh \theta_B}{(\cosh \theta_B - \cos \theta_F)^2}.
\ee
Note that this Jacobian depends only on the polar angles $\theta_B$ and $\theta_F$. It can be verified by direct computation that, with the gauge choice (\ref{Tpar}), the Hamiltonian (\ref{H}) leaves the zero angular momentum sector (relative to $K$ rotations) invariant. This means that acting with the Hamiltonian on a $K$-invariant function $\psi(K^{-1} Q K) = \psi(Q)$, we get another $K$-invariant function. The Hamiltonian (\ref{H}) projected on to the space of $K$-invariant functions has the form (8). It is worth noting that the gauge choice (\ref{Tpar}) is the only gauge choice which leads to a Hamiltonian that preserves angular momentum while being non-singular at the ``north pole'' $\theta_F = \theta_B = 0$.

\section{Spectral Expansion}

\subsection{Eigenfunctions}

To obtain the eigenfunctions of the transfer matrix Hamiltonian (8), the variables $\theta_B$ and $\theta_F$ are decoupled with the help of the Sutherland transformation \cite{SSutherland72}. The resulting decoupled Hamiltonian has the form
\begin{subequations}
\begin{gather}
\tilde{\mathcal{H}} = J^{1/2} \mathcal{H} J^{-1/2} = \tilde{\mathcal{H}}_F + \tilde{\mathcal{H}}_B, \\
\tilde{\mathcal{H}}_F = -\frac{\partial^2}{\partial \theta_F^2} - \frac{1}{4 \sin^2 \theta_F} + \frac{m^2}{4} \frac{1}{\cos^2 (\theta_F/2)}, \\
\tilde{\mathcal{H}}_B = -\frac{\partial^2}{\partial \theta_B^2} - \frac{1}{4 \sinh^2 \theta_B} - \frac{m^2}{4} \frac{1}{\cosh^2 (\theta_B/2)}.
\end{gather}
\label{HBF}
\end{subequations}
The spectrum of the operator $\tilde{\Hc}_F$ (compact sector) is discrete. The eigenfunctions and eigenvalues are given by
\begin{subequations}
\label{Eigenfunctions}
\begin{equation}
 \tilde \phi_l(\theta_F)
  = \sqrt{\sin \theta_F} \cos^m(\theta_F/2) P^{(0,m)}_{(l - m - 1)/2}(\cos \theta_F),
 \qquad
 \epsilon_l
  = \frac{l^2}{4},
\end{equation}
with $l \in 2\mathbb{N} + m + 1$ and $P^{(\alpha,\beta)}_{\nu}(x)$ the Jacobi polynomial of order $\nu$. The values of the discrete parameter $l$ are chosen such that the index of the Jacobi polynomial is $0, 1, 2, \ldots$ Notice that all the eigenfunctions have the prefactor $\cos^m(\theta_F/2)$, hence, they vanish at the south pole $\theta_B = \pi$ where the potential is singular. 

The operator $\tilde{\Hc}_B$ (non-compact sector) has both continuous positive and discrete negative eigenvalues, since the potential is attractive and can host bound states. Eigenstates of the continuous spectrum are
\begin{equation}
 \tilde \phi_r(\theta_B)
  = \frac{\sqrt{\sinh\theta_B}}{\cosh^m (\theta_B/2)}\, P^{(0,-m)}_{(i r + m - 1)/2}(\cosh\theta_B),
 \qquad
 \epsilon_r
  = \frac{r^2}{4}
 \label{phir}
\end{equation}
\end{subequations}
with any real $r \geq 0$. Here we use a generalization of the Jacobi polynomial to complex values of the order. It can be written explicitly in terms of the hypergeometric function as
\begin{equation}
 P^{(0,-m)}_{(i r + m - 1)/2}(x)
  = F\left[ \frac{1 - m - ir}{2}, \frac{1 - m + i r}{2}; 1; \frac{1 - x}{2} \right].
\end{equation}

Discrete negative eigenvalues of $\tilde{\mathcal{H}}_B$ correspond to positive imaginary values of $r$ such that the index in Eq.\ (\ref{phir}) is a positive integer. Explicitly, $r = i(m + 1 - 2k)$ with $k = 1, 2, \ldots, \lfloor m/2 \rfloor$.

Normalization of the eigenfunctions (\ref{Eigenfunctions}) follows from the properties of the Jacobi polynomials and hypergeometric function:
\begin{equation}
 \int_0^\pi d\theta_F\, \tilde\phi_l(\theta_F) \tilde\phi_{l'}(\theta_F)
  = \frac{2}{l}\, \delta_{ll'},
 \qquad
 \int_0^\infty d\theta_B\, \tilde\phi_r(\theta_B) \tilde\phi_{r'}(\theta_B)
  = \frac{4 \delta(r - r')}{r \tanh\frac{\pi}{2} (r + i m)}.
 \label{norm}
\end{equation}

The eigenfunctions of the original transfer matrix Hamiltonian $ \Hc$, Eq.\ (8), are given by
\begin{equation}
 \phi_{l,r} (\theta_F, \theta_B)
  = J^{-1/2} \tilde\phi_l(\theta_F) \tilde\phi_r(\theta_B),
 \qquad
 \Hc \phi_{l,r}
  = \frac{l^2 + r^2}{4}\, \phi_{l,r}.
\end{equation}
Their normalization follows directly from Eq.\ (\ref{norm}),
\begin{equation}
 \int d\theta_F\, d\theta_B\, J\, \phi_{l,r}\, \phi_{l',r'}
  = \frac{8 \delta_{ll'} \delta(r - r')}{l r \tanh\frac{\pi}{2} (r + i m)}.
\end{equation}
All the above eigenfunctions vanish at the origin $\theta_B = \theta_F = 0$ since the weight $J$ is singular at this point. In addition, the original Hamiltonian $\mathcal{H}$ possesses a specific zero eigenstate, $\mathcal{H} \phi_0 = 0$ with
\begin{equation}
 \phi_0
  = \frac{\cos^m(\theta_F/2)}{\cosh^m(\theta_B/2)},
\end{equation}
which is missed by the Sutherland transformation and takes the value $1$ at the origin.

\subsection{Evolution operator}

The evolution operator for the Hamiltonian (8) can be written as an expansion (9) in the eigenstates (\ref{Eigenfunctions}). This expansion involves both a sum over $l$ enumerating discrete states in the compact sector and an integral over the continuous variable $r$ labeling the eigenstates in the non-compact sector.
\begin{equation}
 \psi(\theta_B, \theta_F, x)
  = \phi_0 + \sum_{l} \int dr\, c_{l,r}\, \phi_{l,r}\, e^{-(l^2 + r^2)(x/4\xi)}.
 \label{evolution}
\end{equation}
For a moment, we disregard the discrete part of the spectrum in the non-compact sector.

At the start of the evolution, $x = 0$, the function $\psi(\theta_B, \theta_F, 0)$ should vanish everywhere except the north pole $\theta_B = \theta_F = 0$ where it takes the value $1$. This is just an explicit form of the delta function on the supermanifold. The requirement $\psi(0,0,x) = 1$ fixes the unit prefactor of $\phi_0$ in Eq.\ (\ref{evolution}) since all the nonzero eigenfunctions vanish at the north pole. To establish the values of the coefficients $c_{l,r}$, we use the normalization of the eigenfunctions $\phi_{l,r}$, Eq.\ (\ref{norm}). We set $x = 0$, multiply Eq.\ (\ref{evolution}) by $J \phi_{l,r}$, and integrate over $\theta_{F,B}$. The result of such integration should be zero, which allows us to express the coefficient $c_{l,r}$ as
\begin{equation}
 c_{l,r}
  = -\frac{lr}{8}\, \tanh\frac{\pi}{2} (r + i m) \int d\theta_F\, d\theta_B\, J\, \phi_{l,r}\, \phi_0.
 \label{clr}
\end{equation}

The functions $\phi_0$ and $\phi_{l,r}$ are different eigenfunctions of the Hamiltonian $\mathcal{H}$. Using this fact, we can compute their overlap using the Wronskian method. Define the vector $\mathbf{W}$ using the density flow operator:
\begin{equation}
 \mathbf{W}
  = J \left[ \phi_0 \frac{\partial}{\partial\vec\theta}\; \phi_{l,r} - \phi_{l,r} \frac{\partial}{\partial\vec\theta}\; \phi_0 \right].
 \label{W}
\end{equation}
Here we use the two-dimensional vector notation $\vec\theta = \{\theta_F, \theta_B\}$. Divergence of the vector $\mathbf{W}$ is related to the action of $\mathcal{H}$ on the two eigenfunctions:
\begin{equation}
 \mathop{\mathrm{div}} \mathbf{W}
  = -J \Bigl[ \phi_0 \mathcal{H} \phi_{l,r} - \phi_{l,r} \mathcal{H} \phi_0 \Bigr]
  = -\frac{l^2 + r^2}{4}\; J\, \phi_{l,r}\, \phi_0.
 \label{divW}
\end{equation}
This is exactly the integrand in Eq.\ (\ref{clr}).

Upon substitution of Eq.\ (\ref{divW}) into Eq.\ (\ref{clr}) and applying the Gauss theorem, we can reduce the expression for $c_{l,r}$ to a one-dimensional surface integral. The normal component of the flow vector $\mathbf{W}$ vanishes at the boundaries of the integration domain in $\theta_{F,B}$. This suggests that the overlap integral is zero as a consequence of the orthogonality of the eigenfunctions $\phi_0$ and $\phi_{l,r}$. However, the integration weight $J$ is singular at the north pole $\theta_B = \theta_F = 0$ hence the vicinity of this point should be studied accurately. In order to do this, we will exclude a small circular region of the radius $a$ in the $\theta_B$-$\theta_F$ plane from the integral in Eq.\ (\ref{clr}). At the boundary of this region $\theta_F = a \cos\chi$ and $\theta_B = a \sin\chi$. Expanding $\mathbf{W}$ in $\theta_{F,B} \ll 1$ we obtain
\begin{equation}
 \mathbf{W}
  \approx \left\{ \frac{4 \theta_B \theta_F^2}{(\theta_B^2 + \theta_F^2)^2},\;\; \frac{4 \theta_B^2 \theta_F}{(\theta_B^2 + \theta_F^2)^2} \right\}
  = \frac{4}{a} \Bigl\{ \sin\chi \cos^2\chi,\;\; \sin^2\chi \cos\chi \Bigr\}.
\end{equation}
We integrate the flow of $\mathbf{W}$ through this small circle yielding
\begin{equation}
 \int d\theta_F\, d\theta_B\, \mathop{\mathrm{div}} \mathbf{W}
  = -\int_0^{\pi/2} a\, d\chi\; \mathbf{W} \cdot \{ \cos\chi,\;\; \sin\chi \}
  = -2.
 \label{flow}
\end{equation}
This result is independent of $a$ and remains valid in the limit $a \to 0$.

Using Eqs.\ (\ref{clr}), (\ref{divW}), and (\ref{flow}), we finally obtain
\begin{equation}
 c_{l,r}
  = -\frac{lr}{l^2 + r^2}\, \tanh\frac{\pi}{2} (r + i m).
\end{equation}

The full explicit form of the evolution operator (\ref{evolution}) becomes
\begin{multline}
 \psi(\theta_B, \theta_F, x)
  = \frac{\cos^m(\theta_F/2)}{\cosh^m(\theta_B/2)} \Biggl[
     1 - (\cosh\theta_B - \cos\theta_F) \sum_{l} \int_{im - \infty}^{im + \infty} dr\, \frac{2lr \tanh\frac{\pi}{2} (r + i m)}{2(l^2 + r^2)} \\
     \times P^{(0,m)}_{(l - m - 1)/2}(\cos \theta_F) P^{(0,-m)}_{(i r + m - 1)/2}(\cosh\theta_B)\, e^{-(l^2 + r^2)(x/4\xi)}
    \Biggr]. 
\end{multline}
Here we have extended the $r$ integration to the full real axis (since the integrand is an even function) and raised the contour in the complex plane by $im$. This choice of the integration contour picks a set of poles of $c_{l,r}$ on the imaginary $r$ axis and thus automatically includes the contribution of the discrete levels in the non-compact sector.


\begin{thebibliography}{99}
\bibitem{footnote1}
In relatively narrow WSM samples, surface states may provide a significant contribution to transport. This is particularly relevant when the Landau level degeneracy $m$ is of order one. However, such a contribution can be avoided if the magnetic field is oriented parallel to the separation between Weyl nodes of opposite chirality.
\bibitem{footnote2}
Both internode and intranode scattering rates enter our formalism as parameters and they may, in principle, depend on magnetic field. However such dependence will be most pronounced in strong enough fields close to the clean ultraquantum limit and will not be important for the fields we are considering, when many Landau levels intersect the Fermi surface.
\end{thebibliography}
\end{document}